# Electromagnetic Feature Extraction in Superconducting Quantum Circuits

## An Open-Source Finite-Element Workflow Using Palace

Jiale Ye
Weixian College
Tsinghua University
Beijing, China
yejl22@mails.tsinghua.edu.cn

Jiaheng Wang
School of Integrated Circuits
Tsinghua University
Beijing, China
wangjiah21@mails.tsinghua.edu.cn

Yu-xi Liu
School of Integrated Circuits
Tsinghua University
Beijing, China
yuxiliu@tsinghua.edu.cn

*Abstract*—Accurate electromagnetic (EM) feature extraction, including element characterization, eigenmodes, and field distributions, is essential for superconducting quantum circuit design. To streamline this process, we present a workflow built around Palace, an open-source, high-performance finite element method solver tailored for quantum applications. Starting from circuit layouts, the workflow automates mesh generation, multiple EM solver processing, and EM-to-Hamiltonian post-processing. We benchmark the workflow on a chip with bare resonators and qubits coupled with readout resonators, achieving resonator frequencies prediction within 0.3% and 3 out of 4 external couplings within 16% of cryogenic measurements. These results demonstrate open-source EM tools can match commercial accuracy while offering scalable, license-free analysis. Our Palace-based workflow provides an accessible and extensible foundation for rapid superconducting circuit development and materials-loss studies.

## I. Introduction

Quantum-information hardware spans a range of platforms, including trapped ions [1], photonics [2], spin-based systems [3], topological qubits [4], with superconducting integrated circuits emerging as the most industrially mature platform due to their compatibility with microwave control, scalable lithographic fabrication, and long coherence time at millikelvin temperatures [5].

Realizing these advantages in practice hinges on high-fidelity electromagnetic (EM) feature extraction—such as impedance matrices, port responses, and eigenmode distributions—to convert lithographic layouts into accurate quantum Hamiltonians. These Hamiltonians define the energy landscape of the system and underpin both quantum algorithm execution and experimental verification. In addition, EM simulation is critical for identifying and mitigating non-idealities such as parasitic coupling, mode crowding, and radiative loss [6].

Commercial finite element platforms such as SONNET, COMSOL, and Ansys HFSS provide this simulation capability, but their high license costs and limited cloud scalability can impede iterative design [7]. Moreover, they lack native support for quantum-specific metrics, including energy participation ratio (EPR) [8] and interface-based surface loss estimation, often requiring researchers to implement and maintain custom post-processing tools.

To overcome these constraints, we adopt Palace, an open-source, finite element method (FEM) solver built on robust open-source libraries including MFEM, libCEED and linear system solvers like Hypre, SuperLU. Palace integrates arbitrary high-order finite element spaces and curvilinear mesh support, adaptive mesh refinement, adaptive fast frequency sweep, built-in EPR and interface participation calculators, delivering exascale-ready performance on commodity cloud nodes without license cost.

This paper contributes a streamlined Palace-centered workflow that starts from GDSII format layouts, automates meshing, eigen, and scattering parameter (S-parameters) solves, and performs EM-to-Hamiltonian conversion. We integrated the workflow into an open-source library, PalaceForCQED. We validate the workflow on a transmon (a physical implementation of qubit) chip with dedicated readout resonators, where simulated resonator frequencies agree with cryogenic measurements to within 0.3 %, 3 out of 4 external couplings within 16%, demonstrating that open-source EM extraction can match commercial accuracy while scaling efficiently to larger devices. The resulting workflow furnishes an accessible foundation for rapid superconducting circuit optimization and systematic materials-loss studies.

## II. Quantum Circuits and Electromagnetic Problems

For superconducting quantum computation, a qubit realized by circuits is an anharmonic LC oscillator formed by a Josephson junction shunted by linear capacitance and inductance; variants such as the transmon or fluxonium differ mainly in junction energy and shunt geometry [9] and we normally incorporate coplanar waveguide (CPW) resonators for readout, control, and qubit–qubit coupling in the 5-15 GHz band [6].

The design of such circuits involves translating an abstract circuit graph—encoding the desired quantum Hamiltonian—into a physical layout that precisely reproduces the target Hamiltonian, as the system's quantum dynamics are entirely governed by its Hamiltonian. This requires careful optimization of circuit parameters, including Josephson junction energies, capacitive couplings, and inductive elements, to ensure the emergent low-energy physics matches the intended quantum model. However, unlike idealized lumped-element models, physical implementations introduce complexities such as parasitic couplings, radiative losses, and distributed effects, which can degrade coherence and induce crosstalk. EM simulations enable designers to predict key features directly from geometry, which include static reactance matrices for circuit modelling, eigen-frequencies and EPR calculations for each relevant mode to quantify qubit anharmonicity and loss, and port responses to set resonator frequencies and external couplings.

These features make it possible to build a circuit model from the physical circuits and apply the well-developed methods like Lumped Oscillator model [11], EPR quantization [8], and Black-Box Quantization to obtain the circuit Hamiltonian [12], [13].

To conclude, designing superconducting circuits for experiments reduces to one central task: extracting EM features from a physical layout with sufficient accuracy. These requirements mainly correspond to three types of problems:

- Electrostatic/magnetostatic analysis: extract capacitance and inductance matrices of superconducting circuits.
- Eigenmode analysis: find the characteristic resonant modes and EM field distribution within circuits.
- Frequency-domain driven analysis: obtain the port scattering parameters from circuit input/output ports.

To solve the problems, we adopt Palace, which provides solvers for all these EM problems together with some additional convenience features: the built-in support for EPRs and adaptive fast frequency sweep algorithms for frequency domain driven simulation.

## III. METHOD

As shown in Fig.1. , the EM extraction workflow follows four stages: (i) meshing, (ii) parameters and solver setup, (iii) solver execution, and (iv) data post-processing. All are driven by our library PalaceForCQED together with Palace. A single GDSII file with configurations for meshing and solvers can feed the entire chain, eliminating the multi-tool scripting overhead common in commercial workflows. The workflows are detailed below.

### A. Model Meshing

1) Layout import. Superconducting circuits are usually drafted in GDSII files. We convert GDSII layers to Gmsh geometry descriptions with libGDSII, and post-process them for better mesh quality around curves.
2) Mesh construction. We interface with the Gmsh API to generate structured meshes with configurable parameters, adding ports, substrate, metal layer, and air domain. Mesh sizes are refined near critical regions like CPW gaps. Palace also accepts other mesh formats like VTK, NASTRAN (.nas, .bdf), and COMSOL (.mphtxt, .mphbin) to seamlessly integrate with existing CAD workflows.

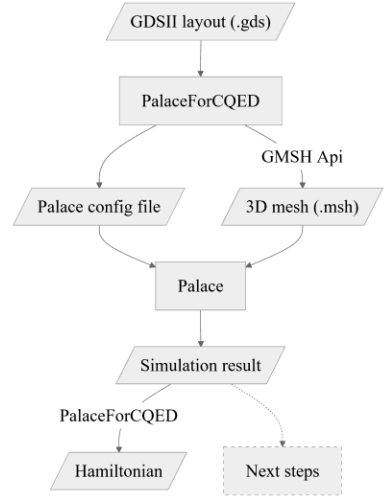

Fig.1. PalaceForCQED workflow for electromagnetic extraction and Hamiltonian construction.

### B. Parameters Setup

Metals are treated as perfect electric conductors, a valid approximation above the superconducting gap. Float-zone Silicon is modelled with $\varepsilon_r = 11.49$ and a cryogenic loss tangent $\tan \delta = 2.3 \times 10^{-6}$ for the results shown in this paper [14]. Lumped-ports on the CPW feed are normalized to 50 Ω to match the room-temperature readout line.

Each Josephson junction is replaced by a lumped inductor $L_J = \Phi_0/2\pi I_c$, where $I_c$ is the junction's critical current [15]. This linear model reproduces EM participation accurately while relegating non-linearity to the quantum-post-processing stage.

### C. Solver Execution

1) Electrostatic pass. Calculate the mutual capacitance matrix for the lumped circuit model.
2) Eigenmode pass. Calculate eigenfrequencies and field maps. Palace automatically outputs EPRs when lumped-port tags coincide with the linearized junctions.
3) Driven-frequency pass. Scattering parameters within the interested frequency ranges using Palace's adaptive fast-frequency sweep with lumped port excitation.

### D. Post-Processing

Simulation outputs, like eigenmodes, capacitance matrices, and S-parameters, are parsed by PalaceForCQED, which performs unit conversions, scaling, and mode filtering. The data is then used to build a quantized circuit Hamiltonian suitable for numerical analysis or pulse-level control simulation.

## IV. RESULTS

Accurate extraction of eigenfrequencies and mode fields in superconducting quantum circuits requires careful tuning of mesh resolution and finite element order. In full-wave FEM, higher-order elements improve convergence per degree of freedom (DoF), but at the cost of increased computational complexity. Striking the right balance between accuracy and computational cost is therefore a critical consideration in electromagnetic design workflows.

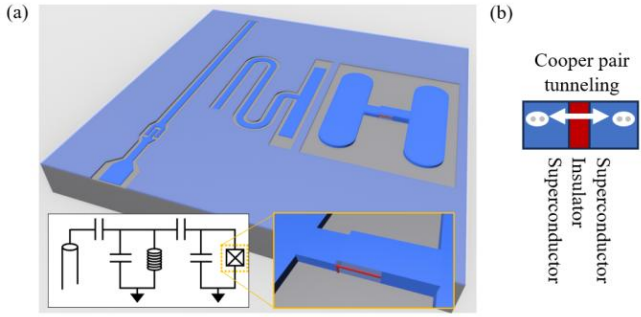

Fig.2. (a) Diagram of a typical transmon qubit-readout resonator system (1Q–1R). The transmon qubit comprises two planar capacitor pads connected by a Josephson junction. The qubit is capacitively coupled to a quarter-wave coplanar waveguide (CPW) resonator, which is itself weakly coupled to a CPW feedline for external drive and readout. The inset shows the equivalent lumped-element circuit schematic of the system. (b) Schematic diagram of a Josephson junction. A Josephson junction is a non-linear device composed of two superconductors separated by a thin insulating barrier, allowing for the tunneling of Cooper pairs.

We performed a convergence study using Palace's eigenmode solver on a CPW resonator. Mesh density was controlled by a dimensionless ratio r, defined via the minimum mesh size $S_{\min} = 1.5 \cdot W_{\text{tr}} \cdot 2^{-r}$, where $W_{\text{tr}}$ is the center trace width of the CPW. We varied both $r$ and the finite element order, then tracked the eigenfrequencies and associated DoFs. Convergence stabilizes at fourth-order elements with $r = 1.5$ (hereafter "r1.5o4"), yielding eigenfrequencies within ±0.3% of the extrapolated asymptotic values while keeping an acceptable elapsed simulation time, illustrating that this setting achieves an optimal tradeoff between accuracy and efficiency. All subsequent simulations reported in this work use the r1.5o4 configuration.

To evaluate the accuracy of the proposed workflow, we simulated the eigenfrequencies and external coupling strengths of a transmon qubit-readout resonator (in short, 1Q-1R) system as shown in Fig.2. . Then, we compared the results with the measurement on a previously fabricated and characterized superconducting chip. The chip comprises two 1Q-1R systems with resonator R1, R2, and two bare quarter-wave CPW resonators R3, R4. These resonators are positioned at different distances from the main feedline waveguide, resulting in distinct coupling coefficients and slight shifts in eigenfrequencies due to loading effects. Using r1.5o4 hyperparameters, we apply the workflow to extract the base resonator frequencies and external coupling strengths. Results are summarized in Table I.

TABLE I. Comparison of Simulation and Experimental Results

| Resonators | Base Frequency (GHz) | | Coupling Strength (MHz) | |
|---|---|---|---|---|
| | *Simulation* | *Experimental* | *Simulation* | *Experimental* |
| R1 | 7.1933 | 7.1787 | 0.3600 | 0.3130 |
| R2 | 7.2861 | 7.2684 | 0.0960 | 0.0903 |
| R3 | 7.5604 | 7.5484 | 0.0916 | 0.1082 |
| R4 | 7.6551 | 7.6447 | 0.1740 | 0.2312 |

The simulated base frequencies match the experimental values within 0.3% across all four resonators. Coupling strengths, except R4, agree within ±16%, which is consistent with expected variation due to packaging uncertainty and sensitivity to boundary conditions. This demonstrates that open-source tools can reliably replace commercial solvers in superconducting circuit analysis workflows.

## V. Conclusion

We identified EM feature extraction as a core challenge in superconducting quantum circuit design and mapped it to three key simulation tasks: electrostatic, eigenmode, and driven-frequency analysis. To address this, we developed an open-source finite element workflow based on Palace to automate quantum circuits simulation and post processes. Our results show that this open workflow achieves commercial-level accuracy while offering a scalable, license-free foundation for rapid layout iteration and systematic loss analysis in next-generation quantum hardware.

## References

[1] D. Kielpinski, C. Monroe, and D. J. Wineland, "Architecture for a large-scale ion-trap quantum computer," Nature, vol. 417, no. 6890, pp. 709–711, Jun. 2002.

[2] J. L. O'Brien, A. Furusawa, and J. Vučković, "Photonic quantum technologies," Nature Photon, vol. 3, no. 12, pp. 687–695, Dec. 2009.

[3] B. E. Kane, "A silicon-based nuclear spin quantum computer," Nature, vol. 393, no. 6681, pp. 133–137, May 1998.

[4] C. Nayak, S. H. Simon, A. Stern, M. Freedman, and S. Das Sarma, "Non-Abelian anyons and topological quantum computation," Rev. Mod. Phys., vol. 80, no. 3, pp. 1083–1159, Sep. 2008.

[5] M. Kjaergaard et al., "Superconducting Qubits: Current State of Play," Annual Review of Condensed Matter Physics, vol. 11, no. Volume 11, 2020, pp. 369–395, Mar. 2020.

[6] E. M. Levenson-Falk and S. A. Shanto, "A Review of Design Concerns in Superconducting Quantum Circuits," Nov. 25, 2024.

[7] N. Shammah et al., "Open hardware solutions in quantum technology," APL Quantum, vol. 1, no. 1, p. 011501, Mar. 2024.

[8] Z. K. Minev, Z. Leghtas, S. O. Mundhada, L. Christakis, I. M. Pop, and M. H. Devoret, "Energy-participation quantization of Josephson circuits," npj Quantum Inf, vol. 7, no. 1, pp. 1–11, Aug. 2021.

[9] P. Krantz, M. Kjaergaard, F. Yan, T. P. Orlando, S. Gustavsson, and W. D. Oliver, "A Quantum Engineer's Guide to Superconducting Qubits," Applied Physics Reviews, vol. 6, no. 2, p. 021318, Jun. 2019.

[10] A. Blais, A. L. Grimsmo, S. M. Girvin, and A. Wallraff, "Circuit quantum electrodynamics," Rev. Mod. Phys., vol. 93, no. 2, p. 025005, May 2021.

[11] G. Burkard, R. H. Koch, and D. P. DiVincenzo, "Multi-level quantum description of decoherence in superconducting qubits," Phys. Rev. B, vol. 69, no. 6, p. 064503, Feb. 2004.

[12] F. Solgun, D. W. Abraham, and D. P. DiVincenzo, "Blackbox Quantization of Superconducting Circuits using exact Impedance Synthesis," Phys. Rev. B, vol. 90, no. 13, p. 134504, Oct. 2014.

[13] S. E. Nigg et al., "Black-box superconducting circuit quantization," Phys. Rev. Lett., vol. 108, no. 24, p. 240502, Jun. 2012.

[14] M. Checchin, D. Frolov, A. Lunin, A. Grassellino, and A. Romanenko, "Measurement of the Low-Temperature Loss Tangent of High-Resistivity Silicon Using a High-Q Superconducting Resonator," Phys. Rev. Applied, vol. 18, no. 3, p. 034013, Sep. 2022.

[15] M. Tinkham, Introduction to Superconductivity: Second Edition, Second edition. Mineola, NY: Dover Publications, 2004.